\documentclass{aa}
\usepackage{graphicx}
\usepackage{yfonts}
\usepackage{xcolor}
\usepackage{txfonts}
\usepackage{subfigure}
\usepackage{float}
\usepackage{natbib,twoopt}
\usepackage{xspace}
\usepackage{booktabs}
\usepackage{balance}
\usepackage{upgreek}
\usepackage{multirow}
\usepackage[normalem]{ulem}
\usepackage{soul}
\usepackage[colorlinks=true, allcolors=blue]{hyperref}
\usepackage[normalem]{ulem}
\usepackage{array}
\usepackage{makecell}
\newcommand\Tstrut{\rule{0pt}{2.6ex}}

\usepackage[flushleft]{threeparttable}

\def\apjl{ApJ Lett.}
\def\apjs{ApJ Suppl.}

\def\aap{A\&A}

\bibpunct{(}{)}{;}{a}{}{,} 
\DeclareGraphicsExtensions{.eps,.ps,.pdf}

\usepackage{amstext}

\begin{document}

\title{A high-resolution study of the double radio relic system in MACS J1752.0+4440}

\author
{M. Della Chiesa\inst{\ref{1},\ref{2}} \and A. Botteon \inst{\ref{2}} \and A. Bonafede\inst{\ref{1},\ref{2}} \and K. Rajpurohit\inst{\ref{4},\ref{2}} \and V. Cuciti\inst{\ref{2}} \and D. Hoang\inst{\ref{3}} \and R. J. van Weeren\inst{\ref{5}} \and X. Zhang\inst{\ref{6}} \and F. Gastaldello\inst{\ref{7}}}

\offprints{Maicol Della Chiesa \\ \email{maicol.dellachiesa@inaf.it}}

\institute{
  DIFA - Università di Bologna, via Gobetti 93/2, I-40129 Bologna, Italy\label{1}
  \and INAF - IRA, via P. Gobetti 101, I-40129 Bologna, Italy\label{2}
  \and Thüringer Landessternwarte, Sternwarte 5, 07778 Tautenburg, Germany\label{3}
  \and Center for Astrophysics | Harvard \& Smithsonian, 60 Garden Street, Cambridge, MA 02138, USA\label{4}
  \and Leiden Observatory, Leiden University, PO Box 9513, 2300 RA Leiden, The Netherlands\label{5}
  \and Max-Planck-Institut fur Extraterrestrische Physik, Giessenbachstrasse, 85748, Garching, Germany\label{6}
  \and INAF - IASF Milano, via A. Corti 12, 20133 Milano, Italy\label{7}
}

\date{Received --; accepted --}

\abstract
   {Radio relics are diffuse, extended synchrotron sources located at the outskirts of merging galaxy clusters. Their origin has been linked with shock waves injected into the intracluster medium, but the acceleration mechanism at the shock front is still under debate. Some clusters, like MACS J1752.0+4440, host a double relics system, with two relics found on opposite sides with respect to the cluster center.}
   {To investigate the acceleration mechanism that generates radio relics, we study the morpholgical and spectral properties of double relic system in MACS J1752.0+4440 ($z=0.366$).}
   {We present new wideband radio continuum observations made with uGMRT and JVLA, and previously published LOFAR data. We perform a detailed, high-resolution spectral analysis of the double relic system in MACS J1752.0+4440, observing and characterizing substructures, particularly for the brighter relic.}
   {We find a double-peaked surface brightness and spectral index profile for the NE relic and identify a ``bright bar'' substructure. Moreover, we observed surprisingly flat integrated spectral indices for both relics, at $\alpha_{\mathrm{int}}^{\mathrm{NE}} = -0.91 \pm 0.06$ and $\alpha_{\mathrm{int}}^{\mathrm{SW}} = -0.83 \pm 0.05$. We study the spatial variation of the spectral index, observing a coherent trend with the observed substructures. We estimate an injection Mach number of $\mathcal{M}_{\mathrm{NE}} = 3.1^{+0.1}_{-0.1}$ and $\mathcal{M}_{\mathrm{SW}} = 3.2^{+0.1}_{-0.1}$. By performing a spectral curvature analysis for both relics, generating color-color plots and a spectral curvature maps, we observe two ``concave’’ spectra represented by positive spectral curvature, in contrast with particle population ageing models.}
   {The observed properties of the NE relic are not consistent with a simple scenario with a single shock front/injection of particles. Multiple shock surfaces, re-acceleration, and projection effects likely play a role in shaping the morphology of the relic.}


\authorrunning{Maicol Della Chiesa et al.}

\titlerunning{A high-resolution study of the double radio relic system in MACS J1752.0+4440}

\maketitle

\section{Introduction}
    Merging galaxy clusters can host several kind of diffuse large-scale synchrotron sources which emission is not associated with individual cluster galaxies. These sources can be broadly classified into radio relics and radio halos \citep[see][for a review]{VanWeeren2019}.
    Their presence unveils the existence of weak magnetic fields ($\sim 0.1 - 10\, \mu$G) and relativistic particles, distributed on cluster-scale, as a non-thermal content of the intracluster medium (ICM). Radio relics are Mpc-size synchrotron sources typically observed in the outskirts of galaxy clusters. They often show irregular arc-like morphologies, curved towards cluster center. The radio spectrum of relics is steep, with $\alpha < -1$ (where the flux density $S_\nu \propto \nu^{\alpha}$), and it is often observed to steepen from the leading outer edge of the relic to the cluster center.\\\\
    The origin of radio relics has been strictly linked with the presence of shock waves injected in the ICM during merger events \citep{Roettiger1999}. Even though the exact formation mechanism is not fully understood, it is believed that a small part of the kinetic energy dissipated during a cluster merger is channeled into non-thermal components \citep{Brunetti2014}. The mechanism responsible for the acceleration of particles is believed to be diffusive shock acceleration (DSA) at the merger shocks \citep[see][for a review]{Bykov2019}. However, several open questions about the viability of DSA are still present. The major open question regards the DSA efficiency, which estimated value is untenable to accelerate thermal particles and produce the radio power observed in most relics, considering that radio relics are associated with weak shocks \citep[$M < 3$,][]{Botteon2020}.
    Moreover, the non-detection of $\gamma$-ray emission from galaxy clusters, which is expected as a product of the collisions between accelerated cosmic ray protons and thermal protons \citep{Vazza2014, Vazza2016}, poses a challenge to the DSA model. Finally, the radio spectral indices of some relics are incompatible with DSA predictions and stationary conditions, showing $\alpha > -1$ \citep[i.e.][]{Stuardi2022}.
    Possible solutions for these problems are the presence of fossil relativistic electrons in the ICM, which can be further accelerated by DSA with lower efficiencies \citep{Bonafede2014, Markevitch2005, Kang2016, vanWeeren2017, Inchingolo2022}, or the presence of other acceleration processes \citep{Zimbardo2018, Kang2019}.\\
    The connection between relics and shocks has been confirmed by locating surface brightness distributions and temperature jumps, at the position of radio relics, in X-rays \citep[i.e.][]{Sarazin2013, Shimwell2015, vanWeeren2016, Botteon2016, DiGennaro2019}. Particularly interesting are the so-called double radio relics, in which two relics are found in diametrically opposite sides from the cluster center, and are believed to trace two shock waves created in a binary and nearly head-on merger event \citep{VanWeeren2011}. A modern and broad-band study on an ensemble of double relic galaxy clusters showed that the merger axis of these object is preferentially found near the plane of the sky \citep{Golovich2019b}.\\\\
    High-resolution studies observed filamentary emission in radio relics \citep[i.e.][]{Owen2014, Rajpurohit2020, Rajpurohit2022, deGasperin2022, Chibueze2023}. While not simple to explain, this filamentary emission is expected and reproduced by magneto-hydrodynamical simulations of cluster merger shocks \citep{Vazza2012, Skillman2013, Wittor2019}.
    \citet{Dominguez-Fernandez2021a} showed how three-dimensional magneto-hydrodynamical simulations of merger shocks are able to reproduce radio relics which radio emission is concentrated into threads and filaments in the shock plane. Their findings highlight the key features that connect the observable characteristics of radio relics with the dynamical properties of the upstream ICM. Furthermore, they point out how weak shocks ($\mathcal{M} \approx 2$) are unlikely to reproduce observable radio relics as they are unable to modify the initial state of the pre-shock magnetic field.
    Moreover, \citet{Wittor2023} showed how observed and simulated filaments have similar properties and that their 3D structures are related to the magnetic field and the Mach number distribution.

\section{MACS J1752.0+4440}
    \begin{figure}
        \centering
        \includegraphics[width=0.9\columnwidth]{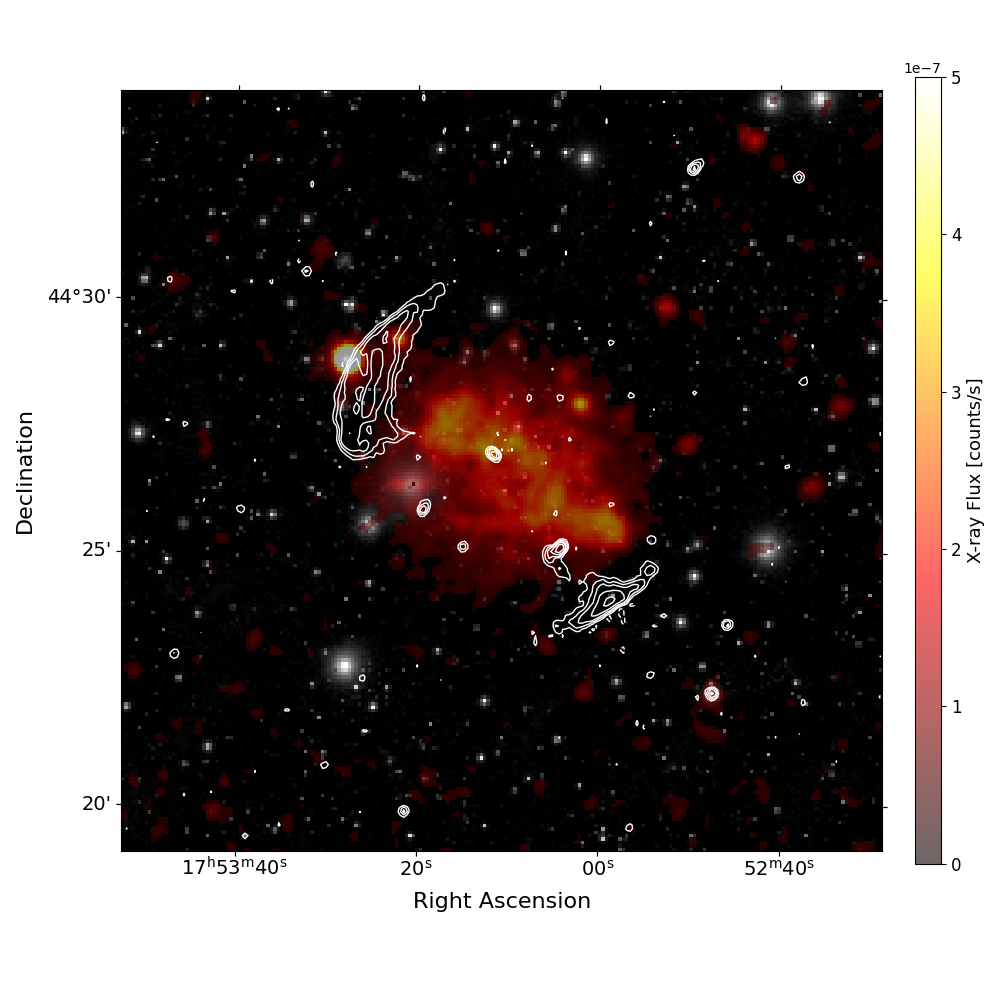}
        \caption{Composite image of MACS J1752.0+4440. The LOFAR 144 MHz contours (in white) are overlaid onto an optical image from Pan-STARRS (in grey) and an X-ray image from XMM-Newton (in red) from \citep{Botteon2022, Zhang2023}.
}
        \label{fig:Opt-X-radio_overlay.png}
    \end{figure}
    The galaxy cluster MACS J1752.0+4440, located at $z = 0.366$, was reported as a candidate host for a double relic system by \citet{Edge2003} as part of the MAssive Cluster Survey \citep[MACS;][]{Ebeling2001}. It was first observed in the radio band in the NRAO VLA Sky Survey \citep[NVSS;][]{Condon1998} and Westerbork Northern Sky Survey \citep[WENSS;][]{Rengelink1997}, which revealed two radio sources found on opposite sides of cluster center.\\
    \citet{vanWeeren2012} studied the cluster in the radio band with WSRT (25, 21, 18, and 13cm) and VLA ($1.4$ GHz) data, detecting a double relic system, with the two relics located North-East (NE) and South-West (SW) of the cluster center, together with a radio halo component. Through an integrated spectral analysis, they obtained spectral indices of $\alpha = -1.16 \pm 0.03$ and $\alpha = -1.10 \pm 0.05$ for the NE and SW relic, respectively. \citet{Bonafede2012} produced spectral index images of the cluster using WSRT (18 cm) and GMRT ($325$ MHz) observations, highlighting a clear steepening of the spectral index for the NE relic, going from the cluster periphery towards cluster center. The SW relic shows a gradient of the spectral index along its main axis. Studying the radio polarization of the two relics, \citet{Bonafede2012} observed that the relics are highly polarized, with polarization values going up to 40\% in the relics' external regions. The high level of polarization suggests an ordered magnetic field at the relics positions. By comparing the observed radio and X-ray morphologies of MACSJ1752.0+4440 with cosmological simulations, they showed that a binary cluster merger with realistic assumptions on shock and turbulence-driven particle acceleration can successfully reproduce the observed features, both in morphology and radio power.
    Finally, MACS J1752.0+4440 belongs to the sample of Planck Sunyaev Zel’dovich detected sources \citep[PSZ2 G071.21+28.86,][]{Planck2016}. These clusters were covered in the  Second Data Release of the LOFAR Two-meter Sky Survey \citep[LoTSS-DR2,][]{Shimwell2022}, analyzed by \citet{Botteon2022}. The analysis of the relics in that sample is reported in \citep{Jones2023}.\\
    
    The optical and X-ray study by \citet{Finner2021} revealed interesting characteristics about MACS J1752.0+4440. By modeling each subcluster with a Navarro-Frenk-White \citep[NFW,][]{Navarro1997} halo they estimate a $1:1$ mass ratio merger, with a total mass of $M_{200} = 14.7_{-3.3}^{+3.8}\times10^{14}\, M_\odot$. In addition, a spectral analysis by \citet{Golovich2019b} suggests an head-on major merger along the plane of the sky, with a time since collision of $0.9 \pm 0.1\,$Gyr. X-ray observations show two clumps of gas at a projected distance of $~1.2\,$Mpc, supporting the dynamical state of the system  (Fig. \ref{fig:Opt-X-radio_overlay.png}). An X-ray surface brightness analysis by \citet{Finner2021} shows two peaks, connected by a bridge of emission, as evidence of a small impact parameter. The two X-ray brightness peaks coincide with cold fronts. However, the available X-ray data were not sufficiently deep to detect a shock front in the cluster's outskirts.\\
    
    In this work, we present the results of new observations of the galaxy cluster MACS J1752.0+4440, combining data from the upgraded Giant Metrewave Radio Telescope (uGMRT) and Karl G. Jansky Very Large Array (JVLA) and data from the LOw Frequency ARray (LOFAR). These observations are able to achieve high-resolution, allowing us to observe and characterize substructures in the radio relics of the galaxy cluster in great details.
    
    The layout of this paper is structured as follows: Section \ref{ObsDataRed} presents a summary of the observations along with details regarding the data reduction process. Section \ref{Results} presents the newly obtained radio images and the associated spectral analysis. Section \ref{Discussion} presents an explanation for the observed findings. Finally, section \ref{Summary} presents a summary of the analysis.
    
    Throughout this paper we assume a $\Lambda$CDM cosmology with $H_0 = 72\, \text{km}\, \text{s}^{-1}\, \text{Mpc}^{-1}$ , $\Omega_m = 0.27$, and $\Omega_\Lambda = 0.73$. At the cluster’s redshift, $1$" corresponds to a physical scale of $4.98$ kpc. All output images are in the J2000 coordinate system and are corrected for primary beam attenuation.

\section{Observations and Data reduction}\label{ObsDataRed}

\subsection{uGMRT}
    MACS J1752+4440 was observed, in June 2021, in band 3 ($300-500\,$MHz) and band 4 ($550-750\,$MHz) for a total of 5 hours in each band, book-ended by ten-minute calibrator scans (3C286). Both band 3 and band 4 observations span a $200\,$MHz frequency range. The dataset for band 3 was split into six $33.3\,$MHz slices, while the dataset for band 4 was split into four $50\,$MHz slices. During the analysis of the individual frequency slices, a significant level of radio frequency interference (RFI) was detected in the lowest-frequency slice of band 3. Due to the severity of the contamination, this slice was excluded from subsequent analysis.\\
    These slices were independently processed using the Source Peeling and Atmospheric Modeling \citep[SPAM;][]{Intema2009} pipeline, which measures the ionospheric phase errors toward the strongest sources in the field of view, allowing to derive direction-dependent gains and fitting a phase screen over the entire field of view. The final analysis was carried out by jointly imaging the slices of each observing band.
\subsection{JVLA}
    The JVLA observations of MACS J1752+4440 were carried out in the JVLA L band ($1-2\,$GHz) in three different configurations: D, C and B, for a total of 15 hours, book-ended by ten-minute calibrators scans. Each observation was executed using eleven spectral windows, divided into 64 channels each, and in full polarization, for a final bandwidth of $600\,$MHz, reduced from the nominal band due to flagging. All four polarizations (RR, RL, LR and LL) were recorded. The data were pre-calibrated performing a self-calibration in \texttt{CASA}, flagging through \texttt{AOFlagger} and proceeding with the standard flux and bandpass calibration. Details about the observations are found in Table \ref{ObservationalOverview}\\

\subsection{LOFAR}
    MACS J1752.0+4440 was observed, as part of the LOFAR Two-meter Sky Survey \citep[LoTSS;][]{Shimwell2017, Shimwell2019, Shimwell2022}, with LOFAR HBA. Data were reduced and calibrated using the LoTSS DR2 pipeline \citet{Tasse2021}, followed by an ``extraction+self-calibration'' scheme \citep{vanWeeren2021}. Here, we use the data calibrated in \citet{Botteon2022} to produce new radio images that we combine with the new uGMRT and JVLA data. A detailed description of the reduction process can be found in \citet{Botteon2022}.
\begin{table*}[t]
    \centering
    \caption{Observations parameters.}
    \begin{tabular}{lllll}
        \toprule    
        & LOFAR & uGMRT Band 3 & uGMRT Band 4 & JVLA \\
        \midrule
        Bandwidth & $48\,$MHz & $200\,$MHz & $200\,$MHz & $1\,$GHz \Tstrut \\
        Central Frequency & $144\,$MHz & $416\,$MHz & $650\,$MHz & $1600\,$MHz\\
        Observation dates & June 16, 2019 & June 14, 2021 & August 1, 2021 & June 19, 2012 \\
        Total on-source time & 8 hr & 5 hr & 5 hr & 15 hr \\
        Configurations &   &   &  & B, C, D\\
        \bottomrule
    \end{tabular}
    \label{ObservationalOverview}
\end{table*}
\begin{figure*}[t]
    \vspace{-1ex}
    \centering
    \includegraphics[width=1\textwidth, trim=50 10 50 100, clip]{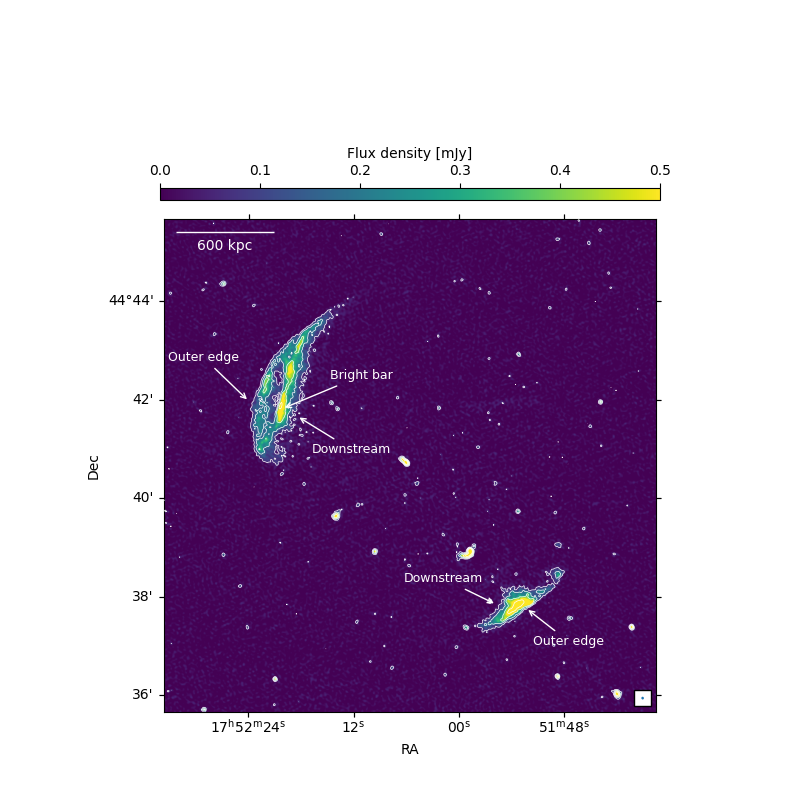}
    \caption{uGMRT Band 4 ($650\,$MHz) image of MACS J1752+4440. This image, at a resolution of $3$''$\times 3$'', is the highest resolution image ever obtained for this cluster. In white, the different structures are labelled. The \textit{outer edge} component outlines the possible position of the merger shock that generated the NE relic, the \textit{bright bar} is a substructure present in the downstream region of the first component. The SW relic shows no substructures at this resolution. Contours are drawn starting at $3\sigma_{rms}$ and are separated by a factor 2. Beam and scalebar are found at the bottom and top of the image, respectively.}
    \label{fig:uGMRT_B4_FullCluster}
\end{figure*}
\subsection{Imaging}
The software used for the deconvolution process was \texttt{WSclean} \citep{Offringa2014}. We produced the high resolution images using a robust weighting of $+0.5$ for the VLA band and $-1$ for every other telescope. We used the multi-scale and multi-frequency deconvolution technique \citep{Offringa2017} to image the dataset, dividing the total bandwidth into narrower frequency channels and interpolating across them to produce a high-fidelity radio image. To speed up and enhance the cleaning process, we produced masks for each image using the \texttt{breizorro} software\footnote{https://github.com/ratt-ru/breizorro}. The threshold for the masks were chosen by running \texttt{breizorro} using different thresholds values, and then selected the one best representing the relics. A summary of the imaging parameter is given in Table \ref{Tab:ImagingParam}. Images were corrected for the primary beam attenuation using a cosine beam shape \citep{Condon2016}.\\

Since spectral analysis relies on comparisons between images at different frequencies, it is crucial that images from different interferometers sample the same range of angular scales. Therefore, we imaged our data with a common lower \textit{uv}-cut at $150\lambda$, corresponding to the shortest well-sampled baseline in the JVLA dataset. This minimizes differences in spatial sensitivity between datasets, ensuring more reliable and self-consistent spectral comparisons. Furthermore, we convolved the images to a common restoring beam of $7''$, forcing a circular beam, using the \texttt{imsmooth} tool in \texttt{CASA}. Finally, we verified the consistency of the integrated flux densities between the original full-resolution maps and the convolved $7''$ images to ensure flux conservation during the smoothing process.\\
\begin{table*}[h]
\centering
    \caption{Imaging parameters.}
    \begin{tabular}{cccccccc}
        \hline
        Frequency  & Restoring beam & Smoothed beam & \# of channels & Briggs & uv-cut & \makecell{\text{RMS  noise} \\ \text{[}$\mu$\text{Jy/beam]}} & \makecell{\text{RMS noise at }$7''$ \\ \text{[}$\mu$\text{Jy/beam]}}\\ \hline
        $144\,$MHz & $6.5$''$\times 4.1$'' & $7''$ & $6$  & $-1$ &$\geq 0.15k\lambda$&  $150$ & $280$\\
        $416\,$MHz & $5.4$''$\times 4.6$'' & $7''$ & $5$  & $-1$ & $\geq 0.15k\lambda$ &  $34$ & $70$\\
        $650\,$MHz & $3.2$''$\times 2.8$'' & $7''$ & $4$  & $-1$ & $\geq 0.15k\lambda$ &  $17$ & $23$\\
        $1.6\,$GHz & $3.2$''$\times 3.0$'' & $7''$ & $3$  & $+0.5$     & $\geq 0.15k\lambda$ &  $10$ & $15$\\ \hline
    \end{tabular}
    \label{Tab:ImagingParam}
\end{table*}
The flux density measurement's uncertainty was estimated as
\begin{equation}
    \Delta S = \sqrt{(\sigma_{rms})^2N_{beam}+(f \cdot S)^2},
    \label{Eq:FluxUncert}
\end{equation}
where $\sigma_{rms}$ is the RMS noise, $N_{beam}$ is the number of beams covered by the source, $f$ is the absolute flux density calibration uncertainty and $S$ is the flux density. The absolute flux density uncertainty was assumed as $10\%$ for LOFAR \citep{Shimwell2022}, $5\%$ for uGMRT Band 3 and Band 4 \citep{Chandra2004}, and $2.5\%$ for JVLA data \citep{Perley2013}.

\section{Results}\label{Results}
\begin{figure*}
    \centering
    \includegraphics[width=0.8\textwidth]{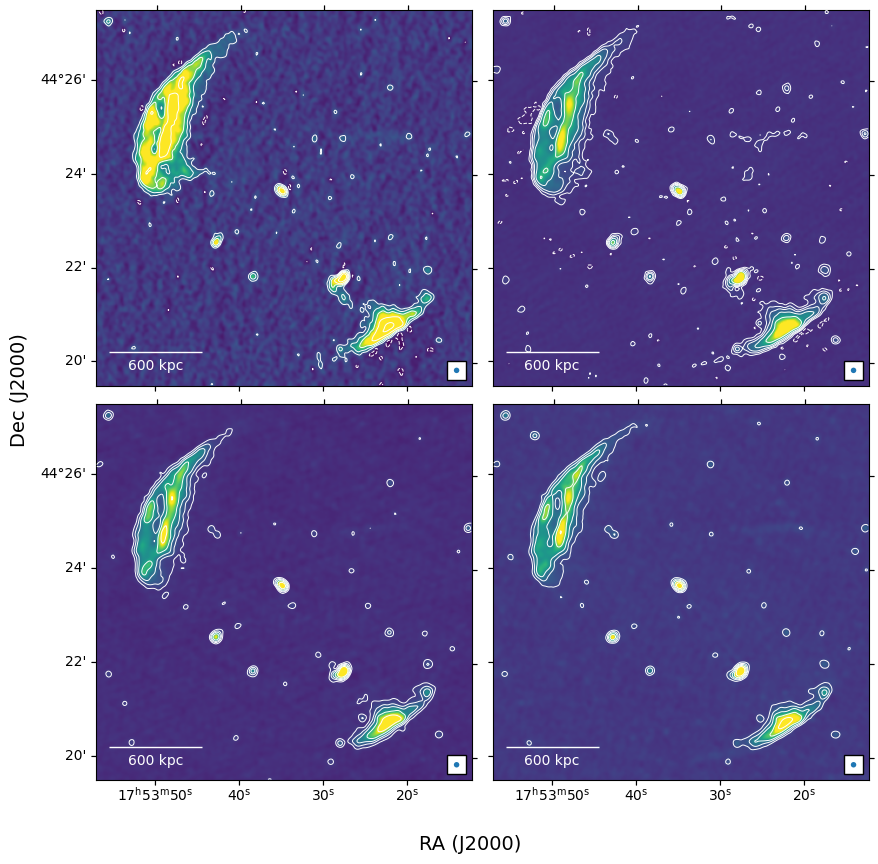}
    \caption{$7$'' resolution radio images for the MACS J1752+4440 cluster. \textit{Top Left}: LOFAR image at $144\,$MHz with noise of $\sigma_{rms,144}=280\,$$\mu$Jy $\text{beam}^{-1}$. \textit{Top Right}: uGMRT image at $416\,$MHz with noise of $\sigma_{rms,416}=70\,$$\mu$Jy $\text{beam}^{-1}$. \textit{Bottom Left}: uGMRT image at $650\,$MHz with noise of $\sigma_{rms,650}=23\,$$\mu$Jy $\text{beam}^{-1}$. \textit{Bottom Right}: JVLA image at $1.6\,$GHz with noise of $\sigma_{rms,1600}=15\,$$\mu$Jy $\text{beam}^{-1}$. Images have a common angular resolution of $7^{\text{"}} \times 7^{\text{"}}$ and contour levels are drawn at $[-3, 3,  9, 18, 36, 72] \times \sigma_{rms}$ (the negative contour is in dashed). The color scale is logarithmic and beam and scalebar are found at the bottom of the images.}
    \label{fig:Cluster_MultiFreq}
\end{figure*}

\begin{figure*}[t]
    \centering
    \begin{minipage}[t]{0.48\textwidth}
        \centering
        \includegraphics[width=0.92\textwidth]{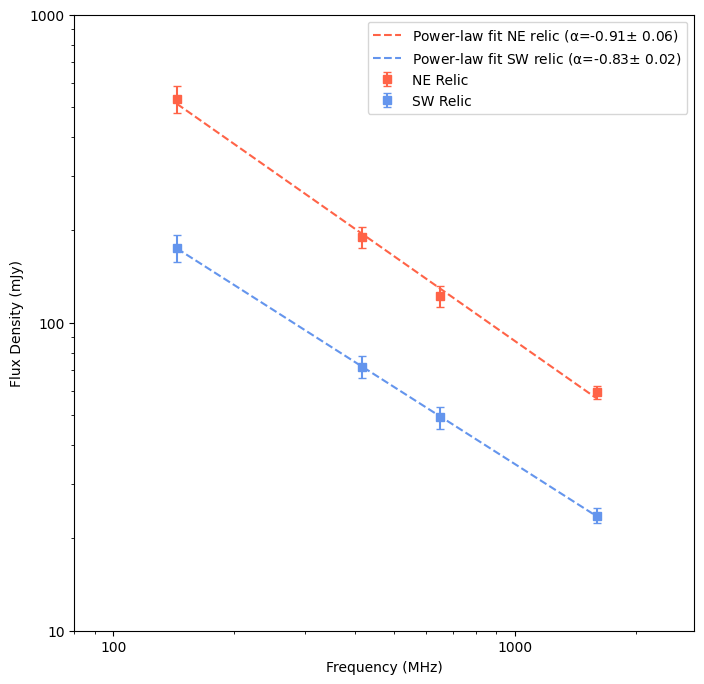}
    \end{minipage}%
    \hfill
    \begin{minipage}[t]{0.48\textwidth}
        \centering
        \includegraphics[width=0.90\textwidth, trim=10 10 10 10, clip]{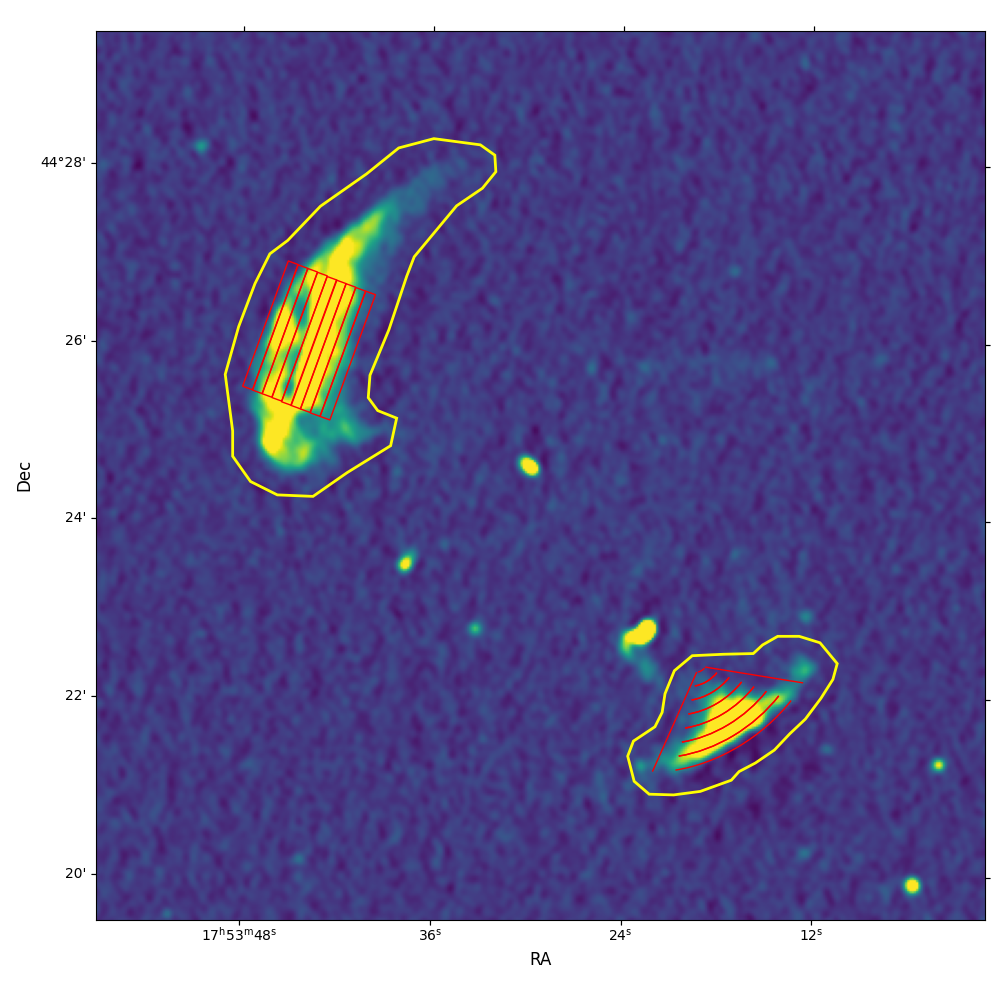}
    \end{minipage}
    \caption{\textit{Left}: Spectra for NE (red) and SW (blue) relic, fitted by two power-laws. \textit{Right}: LOFAR high-resolution continuum image at $144\,$MHz with the regions used to estimate the flux density  of the two relics over-imposed in yellow. Additionally, the regions used to produce the brightness and spectral index profiles are overlaid in red along the relics.}
    \label{fig:IntegFlux_Regions}
\end{figure*}

In Fig. \ref{fig:uGMRT_B4_FullCluster}, we present our deep, $3\text{''}\times3\text{''}$ resolution, uGMRT band 4 image of MACS J1752.0+4440. Thanks to the high resolution of the image, we were able to resolve the fine details of the two radio relics. In particular, the North East (NE) relic presents a substructure along its width. We labelled the NE relic into three components: the ``outer edge'' at the expected position of the shock wave, the ``bright bar'' located in the middle of the relic's width and a ``downstream'' region towards cluster center. The ``outer edge'' and the ``bright bar'' are positioned at $\approx1$ Mpc and $0.9$ Mpc from the cluster center, respectively, where the cluster center was defined as the midpoint of the axis connecting the two relics. The same features can be observed in $7$'' resolution images of each frequency (see Fig. \ref{fig:Cluster_MultiFreq}).\\
The two relics have a largest linear size (LLS) of $\mathcal{L}_{NE} = 1.3\,$Mpc and $\mathcal{L}_{SW} = 0.7\,$Mpc.
We selected two regions enclosing the two radio relics, guided by the LOFAR $3\sigma_{rms}$ contours, at a resolution of $7''$, as shown on the right of Fig. \ref{fig:IntegFlux_Regions}. After checking that no contaminating radio sources were present, we measured the two relics' integrated flux densities, at each frequency. As it can be seen on the left of Fig. \ref{fig:IntegFlux_Regions}, in the considered frequency range, the observed radio spectra appears to be well-represented by a power-law fit. Results are reported in Table \ref{tab:FluxDensities}.
This fit gives two integrated spectral indices $\alpha_{int}^{NE} = -0.91 \pm 0.06$ and $\alpha_{int}^{SW} = -0.83 \pm 0.05$. These values differ from those reported in earlier studies \citep[][]{vanWeeren2012, Bonafede2012}, which were based on lower-resolution and narrower frequency data. The improved resolution of our data allowed us to calculate the spectral index over regions accurately tracing the edges of the radio relics, and the wider frequency range and higher sensitivity allowed us to precisely measure the integrated radio flux densities.

\subsection{Spectral profiles}
\begin{figure*}[t]
    \centering
    \begin{minipage}[t]{0.48\textwidth}
        \centering
        \includegraphics[width=\textwidth]{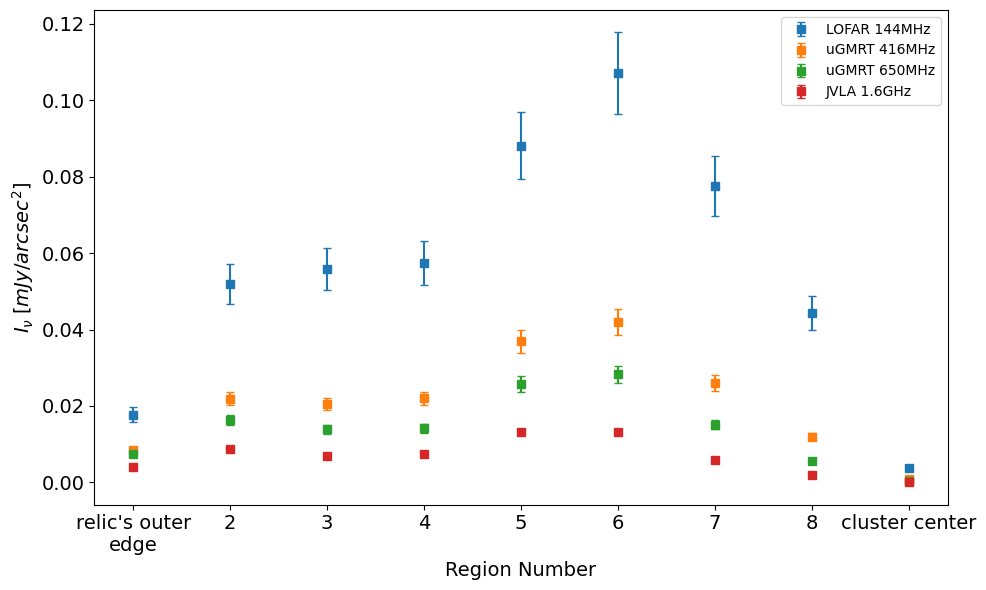}
    \end{minipage}%
    \hfill
    \begin{minipage}[t]{0.48\textwidth}
        \centering
        \includegraphics[width=\textwidth]{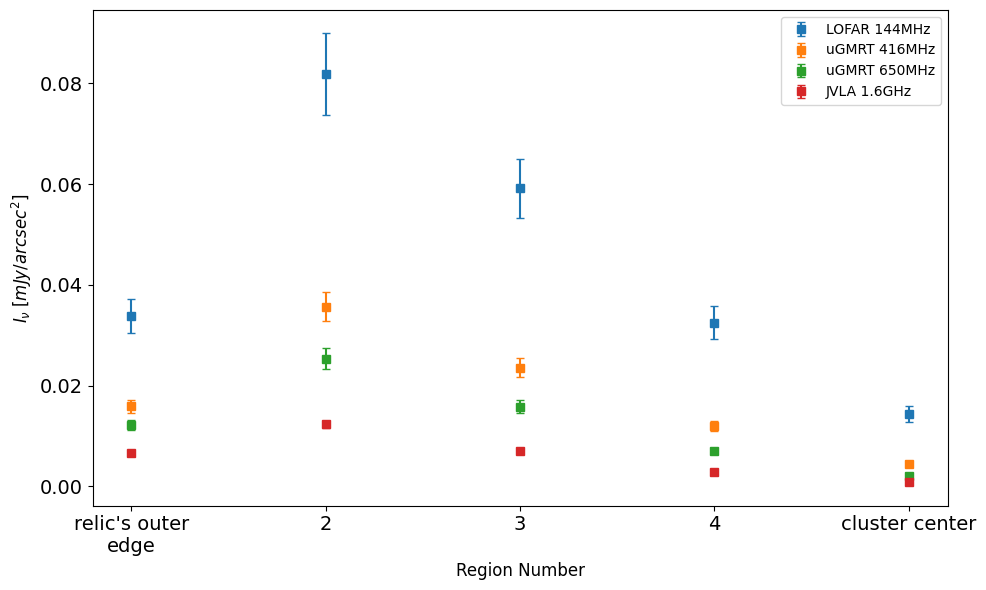}
    \end{minipage}%
    \caption{Surface brightness profiles for both radio relics. \textit{Left}: Profile of the NE relic. The profile presents the peak brightness in region 5 and 6, with a second peak in region 3. This profile is consistent with what can be seen in the spectral index profiles, with two peaks in the position of the two substructures of the relic. \textit{Right}: Profile of the SW relic. The profile shows the peak brightness in region 2, followed by a fast decline in the downstream region.}
    \label{fig:Brightness_profiles}
\end{figure*}

\begin{figure*}[]
    \centering
    \begin{minipage}[t]{0.5\textwidth}
        \centering
        \includegraphics[width=0.9\textwidth]{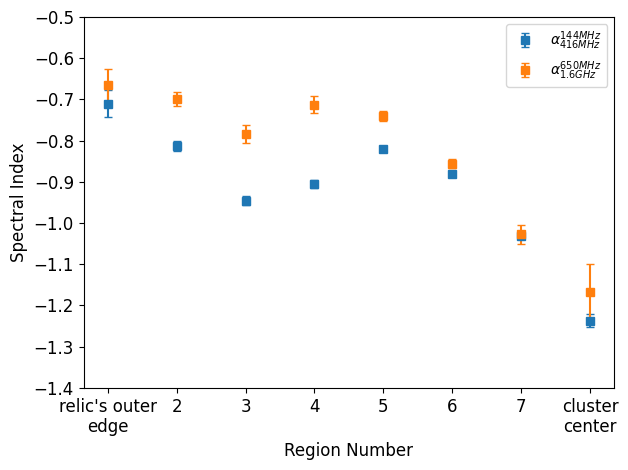}
    \end{minipage}%
    \hfill
    \begin{minipage}[t]{0.5\textwidth}
        \centering
        \includegraphics[width=0.9\textwidth]{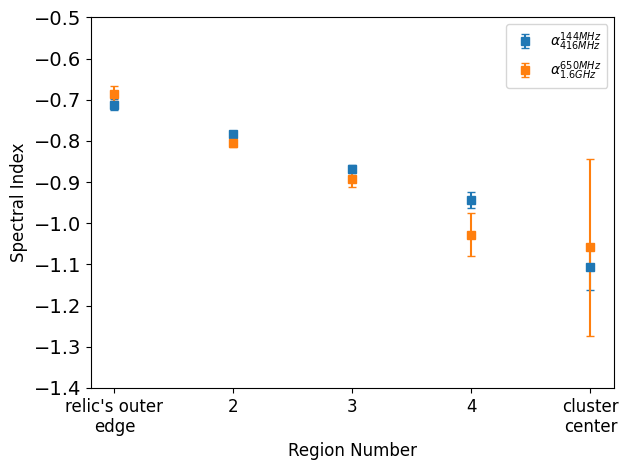}
    \end{minipage}
    \caption{Spectral index profiles, calculated in the two frequency ranges, at high-resolution. \textit{Left}: Profiles for the NE relic. The low-frequency profile shows two peak of spectral index, one at the position of the merger shock and one at the position of the substructure. The profile at high frequency is different, displaying a flat behavior along the regions overlaid to the \textit{bright bar} substructure. Both profiles show a sharp drop of spectral index into downstream regions. \textit{Right}: Profiles for the SW relic. Both profile show a similar trend, decreasing toward cluster center.}
    \label{fig:SpectralIndex_profiles}
\end{figure*}

Given their origin, radio relics typically exhibit synchrotron emission that peaks at the shock front and follows a power-law distribution in both particle energy and radio spectrum \citet{Hoeft2007}. After the shock passage, particle populations are subject to ``ageing'', radiating energy through synchrotron and inverse Compton losses, with higher energy particles (responsible for higher frequency emission) ageing faster, creating a steepening effect in the power-law, at high frequency, in the radio spectrum. In order to compare the properties of the relics in MACS J1752+4440 with the predictions from DSA, we analyzed their surface brightness and spectral index properties.
In the most simple and idealized case, a merger-driven shock, with electrons injected at the shock, would generate a radio relic with a surface brightness profile that peaks at the shock front and decreases towards the cluster center. Numerical simulations of shocks generating radio relics \citep{Dominguez-Fernandez2021, Wittor2023} have shown a more complex scenario. In fact, the increasing resolution of simulations and radio observations \citep{Rajpurohit2020, Chibueze2023} of radio relics have confirmed the presence of substructures and filaments in the downstream of these objects. The ``bright bar'' feature that we observed in the NE relic of MACS J1752 could be a further confirmation of the complex structure of relics.\\
We produced surface brightness and spectral index profiles on regions selected along the relics major axis. These regions were drawn using boxes, with a width of $7''$, equal to the beam size (right panel of Fig. \ref{fig:IntegFlux_Regions}).
Fig. \ref{fig:Brightness_profiles} shows the brightness profile at at $7''$ resolution for both relics. In the NE relic (Fig. \ref{fig:Brightness_profiles}, left), the profile reveals a distinct local maximum that coincides spatially with the 'bright bar' substructure. Beyond this feature, the emission maintains a constant value toward the 'outer edge' of the relic. This profile highlights the presence of two main substructures. On the other hand, the SW relic profile displays the peak brightness in the region behind the position of the ``outer edge''.
Considering an idealized single shock scenario producing particle (re-)acceleration with homogeneous properties of the downstream medium, the expected spectral index profile for a radio relic would be a monotonic decrease of the spectral index, with the flattest spectra at the position of the merger shock, steepening toward the cluster center.\\
To study the origin and infer the spectral properties of the radio emission of the two relics, we produced spectral index profiles at high and low-frequency.\\
The spectral index was computed using the general equation:
\begin{equation}
    \alpha = \frac{\log \left( \frac{S_1}{S_2}\right)}{\log \left( \frac{\nu_1}{\nu_2}\right)},
    \label{SpectralIndex}
\end{equation}
where $S_1$ and $S_2$ are the measured flux densities at $\nu_1$ and $\nu_2$ frequencies.
In Fig. \ref{fig:SpectralIndex_profiles} we compare the spectral indices calculated between LOFAR ($144\,$MHz) and uGMRT Band 3 ($416\,$MHz) along with uGMRT Band 4 ($650\,$MHz) and JVLA ($1.6\,$GHz). The NE relic profiles reveal a non-linear trend, with flattest spectral index coinciding with the position of the ``outer edge'', and a flattening in the central regions. The SW profile shows instead a gradual steepening of the spectral index with no features in the profile. While the ``bright bar'' morphology in the NE relic was visually distinguishable in previous low-resolution studies \citep[i.e.,][]{vanWeeren2012}, it remained unresolved for detailed quantitative analysis. In this work, our higher resolution allow us to move beyond visual inspection. The profiles presented in Fig. \ref{fig:Brightness_profiles} provide a characterization of this substructure's surface brightness and spectral index properties.
\\
\begin{table}[h]
    \centering
    \footnotesize
    \caption{Measured flux densities at each of the four frequencies for the NE and SW relic and integrated spectral index.}
    \begin{tabular}{cccccll}
        \toprule
        \multicolumn{1}{c}{} & \multicolumn{1}{c}{Frequency {[}MHz{]}}                                           & \multicolumn{1}{c}{Flux density {[}mJy{]}}                                                         & $\alpha_{int}$ \\
        \hline
        \multicolumn{1}{c}{NE}        & \multicolumn{1}{c}{\begin{tabular}[c]{@{}c@{}}$144$\\ $416$\\ $650$\\ $1600$\end{tabular}} & \multicolumn{1}{c}{\begin{tabular}[c]{@{}c@{}}$530\pm50$\\ $203\pm20$\\ $130\pm10$\\ $62\pm30$\end{tabular}} &   $-0.91 \pm 0.06$ \\
        \hline
        \multicolumn{1}{c}{SW}        & \multicolumn{1}{c}{\begin{tabular}[c]{@{}c@{}}$144$\\ $416$\\ $650$\\ $1600$\end{tabular}} & \multicolumn{1}{c}{\begin{tabular}[c]{@{}c@{}}$177\pm18$\\ $79\pm6$\\ $56\pm4$\\ $26\pm1$\end{tabular}}      &  $-0.83 \pm 0.05$ \\ 
        \bottomrule
    \end{tabular}
    \label{tab:FluxDensities}
\end{table}
\subsection{Spectral Index map}
\begin{figure}
    \centering
    \includegraphics[width=0.51\textwidth, trim=10 10 0 10, clip]{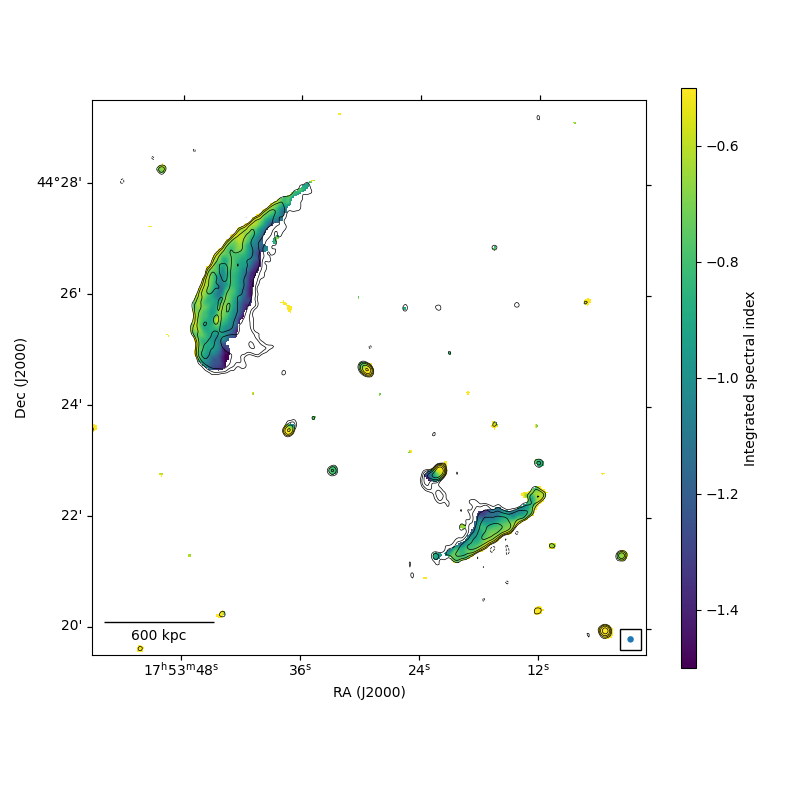}
    \caption{Spectral index map, obtained by fitting the four measurements in each pixel above a $3\sigma_{rms}$ level. Contours are drawn from the LOFAR $144\,$MHz image, starting from a $3\sigma_{rms}$ level and spaced by powers of 2.}
    \label{fig:IntegSpixMaps}
\end{figure}

As we detected multiple components along the NE relic, we investigated the influence of the ``bright bar'' substructure on the spatial variation of the spectral index, across the two relics, by producing the pixel-to-pixel spectral index map (Fig. \ref{fig:IntegSpixMaps}). Across both relics, the spectral index ranges from roughly $\alpha \sim -0.6\; \text{to} -1.4$. Notably, despite the SW relic shows a uniform steepening along its minor axis, the ``bright bar'' substructure in the NE relic is clearly recovered in the spectral index map, where it exhibits flatter values of $\alpha \sim -0.6\; \text{to} -0.8$. This behavior likely contains information on the origin of the relic. This will be discussed in Section 4.

\subsection{Mach numbers}
As previously described, the DSA model for particle acceleration is ordinarily used to describe particle acceleration in astrophysical shocks, which are believed to generate radio relics. In this scenario, considering a power-law energy distribution of relativistic electrons $n_E \propto E^{-\delta}$, the slope $\delta$ is related only to the shock's strength as:
\begin{equation}
    \delta_{inj} = 2\frac{\mathcal{M}^2 + 1}{\mathcal{M}^2 - 1}.
\end{equation}
Under stationary conditions, assuming that the physical conditions in downstream regions do not change significantly with distance from the shock front, the integrated electron spectrum follows a power-law with slope $\delta = \delta_{inj} + 1$. Knowing the relation between the energy distribution and the radio spectral index $\alpha = (1-\delta)/2$, we can directly connect the integrated radio spectral index to the Mach number, as:
\begin{equation}
    \alpha_{int} = -\frac{\mathcal{M}^2 + 1}{\mathcal{M}^2 - 1}.
    \label{integratedalpha}
\end{equation}
In the downstream region, radiative cooling of the electrons leads to a progressively steepening electron spectrum. When considered over a distance exceeding the electron cooling length, the spectral index of the integrated spectrum can be expressed as
\begin{equation}
    \alpha_{inj} = \frac{1}{2} - \frac{\mathcal{M}^2+1}{\mathcal{M}^2-1} = \frac{1}{2} + \alpha_{int}.
    \label{Eq:DSA1}
\end{equation}
As the integrated spectral index for either relics of MACS J1752+4440 is $\alpha_{int} > -1$, the Mach number cannot be calculated through Eq. \ref{integratedalpha}, suggesting a more complex scenario than the one outlined above.\\
However, assuming that we are observing the injected population of electrons, we calculated the Mach number related to the injection spectral index of the shock $\alpha_{inj}$, where the injection spectral index was assumed as the spectral index calculated in the outermost regions, which should trace the position of the shocks. The results obtained for both radio relics and both resolutions are shown in Table \ref{tab:MachNumbers}.
\begin{table}[]
\centering
\caption{Spectral indices and Mach numbers of both radio relics.}
\label{tab:MachNumbers}
    \begin{tabular}{ccccc}
    \hline
       & $\alpha_{inj}$ & $\mathcal{M}_{inj}$ & $\alpha_{int}$ & \\ \hline \\[-6pt]
    NE & $-0.74\pm 0.02$                & $3.1^{+0.1}_{-0.1}$ & $-0.91 \pm 0.06$ & \\[4pt]
    SW & $-0.71 \pm 0.01$               & $3.2^{+0.1}_{-0.1}$ & $-0.83 \pm 0.05$               &  \\[3pt] \hline
    \end{tabular}
    \tablefoot{Integrated Mach numbers cannot be calculated and so are not present.}
\end{table}

\subsection{Spectral curvature}
\begin{figure*}[]
    \centering
    \begin{minipage}[t]{0.5\textwidth}
        \centering
        \includegraphics[scale=0.4, trim=10 10 0 0, clip]{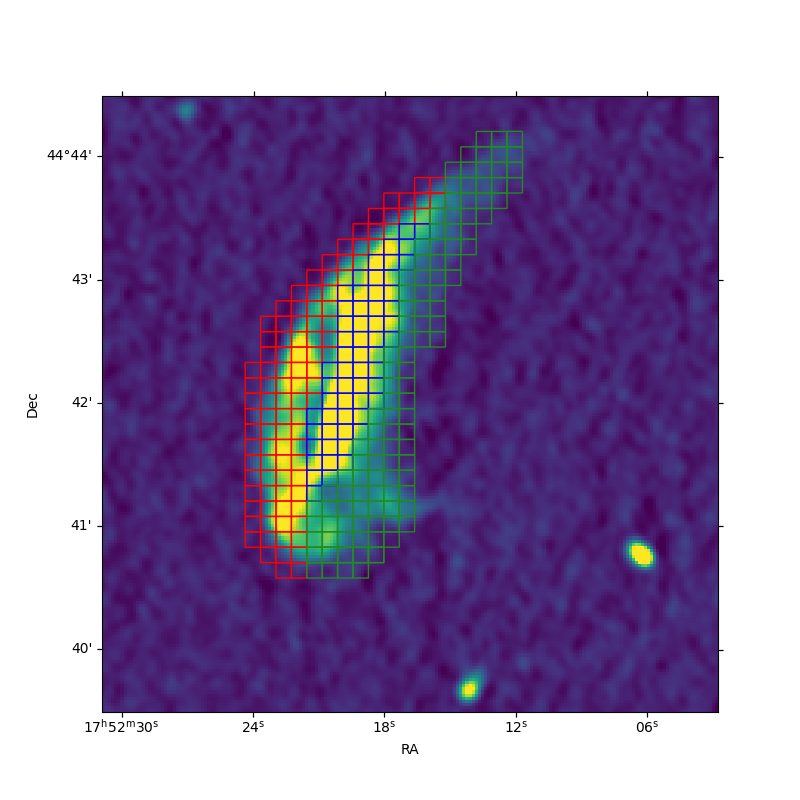}
    \end{minipage}%
    \hfill
    \begin{minipage}[t]{0.5\textwidth}
        \centering
        \includegraphics[scale=0.4, trim=0 10 10 0, clip]{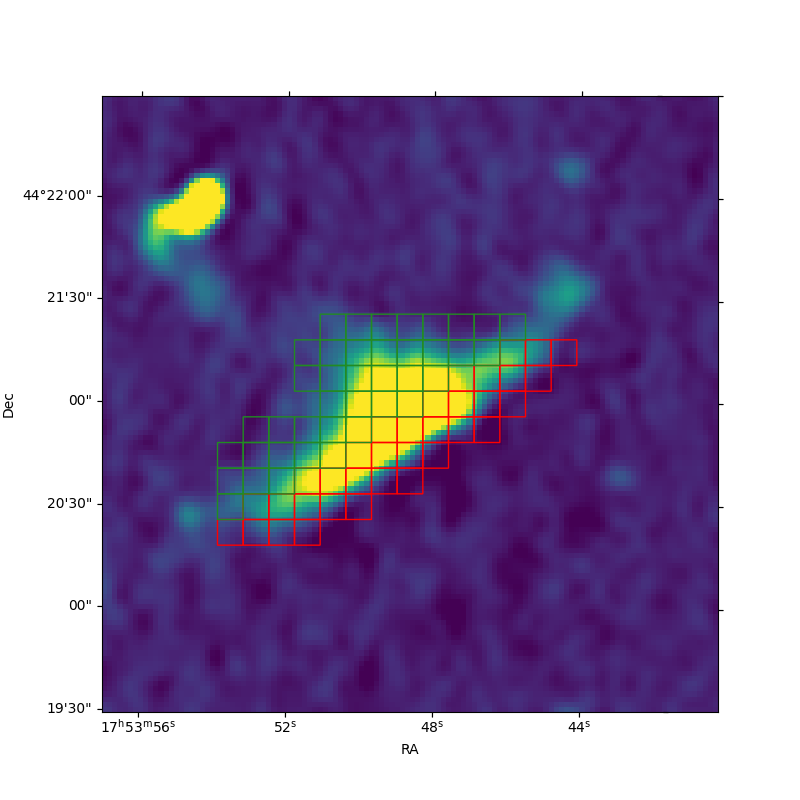}
    \end{minipage}%
    \caption{Zoom in view on the NE and SW radio relic, with selected regions for the curvature study overlaid in different colors. The three components of the NE relic were divided as follows: in \textit{red} we outlined the position of the \textit{outer edge}, in \textit{blue} we followed the \textit{bright bar}, while in \textit{green} we selected the downstream regions. In a similar manner, the shock front and the downstream regions were selected along the SW relic. The same colors are used to represent the structures in the following plots.}
    \label{fig:Color-Color_regions}
\end{figure*}

\begin{figure*}[htbp]
    \centering
    \begin{minipage}[t]{0.5\textwidth}
        \centering
        \includegraphics[width=0.9\textwidth, trim=10 0 0 10, clip]{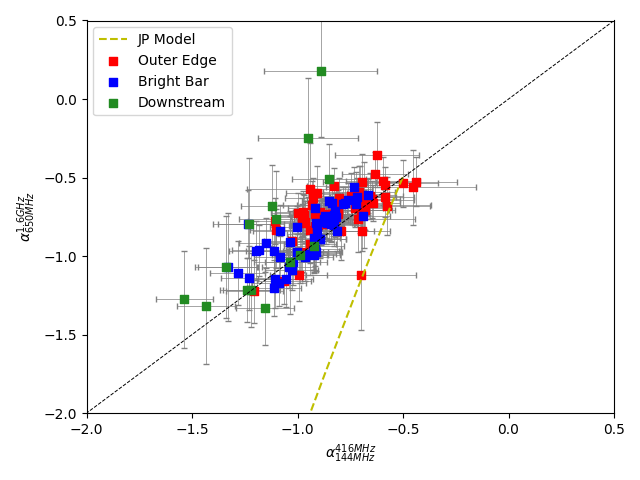}
    \end{minipage}%
    \hfill
    \begin{minipage}[t]{0.5\textwidth}
        \centering
        \includegraphics[width=.99\textwidth, trim=0 0 10 10, clip]{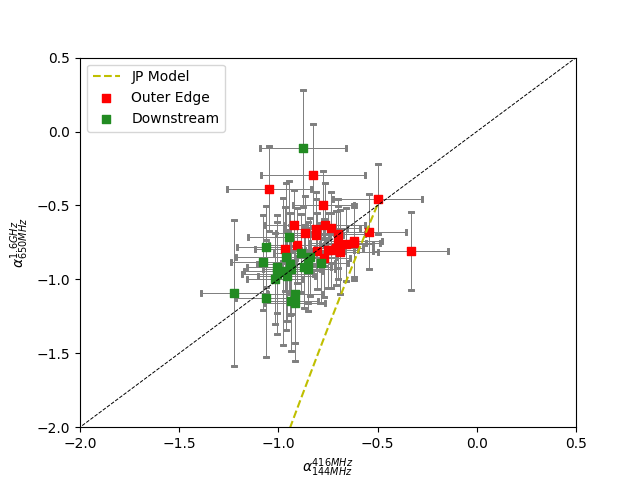}
    \end{minipage}
    \caption{Color-color plot of both radio relics at high-resolution ($7\text{''}$). \textit{Left}: the NE relic shows a lack of points below the power-law line, which indicates a mostly power-law spectrum, with points appearing in the ``concave'' spectrum region. The coloring shows how the \textit{outer edge} and the \textit{bright bar} both fall mostly on the power-law line, with no sign of spectral steepening. \textit{Right}: the SW relic displays some steepening in points at lower spectral index, but still presents multiple points representing a flattening spectrum.}
    \label{fig:Color-Color_plots}
\end{figure*}

As merger-shock models predict increasing spectral curvature in post-shock regions, considering the peculiar structure of the relics of MACS J1752+4440, we have analyzed the curvature of the relics spectra.\\
We employed the color-color diagram technique \citep{Rudnick1994, vanWeeren2012b} which consists in comparing the spectral index calculated between a low (LOFAR $144\,$MHz - uGMRT Band 3 $416\,$MHz) and high (uGMRT Band 4 $650\,$MHz - JVLA $1.6\,$GHz) frequency range, calculated in a grid of boxes of the beam size ($7$") covering the two radio relics (see Fig. \ref{fig:Color-Color_regions}).
Color-color diagrams provide information about the curvature of the spectrum, and thus on particle ageing, by comparing the position of the points with respect to a perfect power-law line (i.e. injection spectrum), which corresponds to $\alpha^{\nu_1}_{\nu_2}=\alpha^{\nu_3}_{\nu_4}$.\\
We considered only the boxes in which the integrated flux density is above $3\sigma_{rms}$ in all images simultaneously. While this guarantees a high signal-to-noise ratio and avoids contamination by noise, it introduces selection biases. Specifically, this threshold preferentially excludes low-surface-brightness regions, which are typically associated with older, more steep-spectrum emission. The color-color plots for both relics are shown in Fig. \ref{fig:Color-Color_plots}.\\
In both relics, and remarkably in the NE relic, we observe points closely following the power-law line. This behavior may be attributed to the presence of multiple factors. For example, fresh or re-accelerated electron populations at multiple sites or the presence of projection effects, where particle populations of different radiative age overlap along the line of sight \citep{Wittor2023}. This behavior suggests that a scenario of a single shock front moving exactly on the plane of the sky is oversimplified. As a simple comparison, a JP model with injection spectrum of $-0.5$ \citep{Jaffe1973} was plotted together with the observed curvature points. The JP model assumes a single burst of particle acceleration and an isotropic angle between the magnetic field and the electron velocity vectors (the so-called pitch angle), on a timescale shorter than the radiative timescale. It is commonly used for post-shock regions, assuming a planar shock front, observed perfectly edge-on \citep{Rajpurohit2020, Rajpurohit2021}. Clearly, the observed curvature points are not represented by the JP model. Other models have been considered in the literature, which take into account a slight inclination along the LoS for the shock front \citep[i.e. KGJP,][]{Komissarov1994}, but as the trend is similar to the one produced by the JP model, we decided to not plot them.

\section{Discussion}\label{Discussion}
\begin{figure}
    \centering
    \includegraphics[scale=0.55]{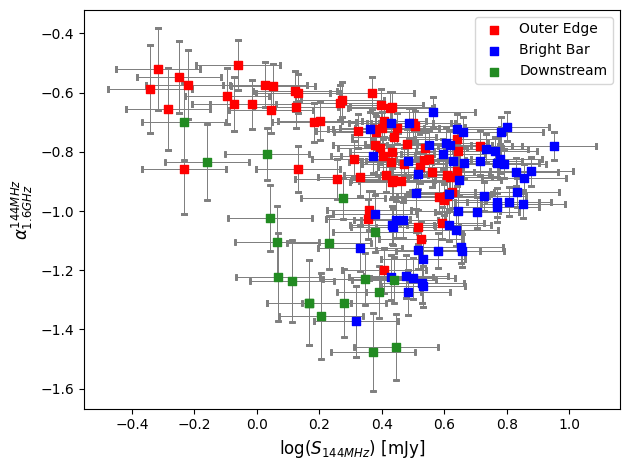}
    \caption{Spectral index against LOFAR $144\,$MHz flux density plot. The spectral index was calculated between LOFAR $144\,$MHz and JVLA $1.6\,$GHz data to leverage on the broadest available frequency range. The two substructures of the NE relic can be recognized in the plot thanks to the high resolution of the images.}
    \label{fig:AlphaFlux}
\end{figure}

The detection of an integrated spectral index flatter than $-1$ in MACS J1752 is highly significant, as it represents a deviation from the typical population of radio relics. Aside from Abell 3365 \citep{Duchesne2021}, MACS J1752 stands as one of the few well-documented cases where the integrated spectrum challenges the fundamental injection limits of standard DSA models. Historically, other candidates have been proposed, but subsequent observations have generally brought them back into alignment with standard DSA predictions. For instance, while \citet{vanWeeren2009} initially reported $\alpha > -1$ for ZwCL 2341.1+0000, recent analysis by \citet{Zhang2021}, using GMRT and JVLA data, yielded significantly steeper indices. Similarly, flat values were initially reported for Abell 3667 \citep{Hindson2014, Riseley2015}. On the contrary, using in-band MeerKAT data, \citet{deGasperin2022} found spectral indexes $<-1$ for both the relics. However, combining the MeerKAT measurement with literature data they found alpha $<-1$ for the NW and $>-1$ for the SE relic.\\
Furthermore, advancements in state-of-the-art observational techniques with progressively increasing resolution have led to more frequent detections of substructures within radio relics, like the one observed in the NE relic of MACS J1752. For example, structures in the form of filaments and twisted ribbons have been recently observed in numerous radio relics \citep{Rajpurohit2020, deGasperin2022, Chibueze2023, Rajpurohit2024}. The physical processes governing the nature of these features are still unclear. In fact, the ``bright bar'' component of the NE relic and the trends of surface brightness and spectral index cannot be explained by an isolated, uniform shock surface.
In order to gain deeper insight into the underlying physical mechanisms, we investigated the potential presence of more than one particle population with different radiative ages along the NE relic structure.
Using the same boxes shown in Fig. \ref{fig:Color-Color_regions}, we plotted the spectral index calculated between LOFAR $144\,$MHz and JVLA $1.6\,$GHz against the LOFAR $144\,$MHz flux density, color coding the points based on the position of the box along the relic. 
In fig. \ref{fig:AlphaFlux}, we observe that the ``outer edge'' spans low to moderate radio fluxes with consistently flat spectra, typical for freshly accelerated electrons. On the other hand, the ``bright bar'' shows higher fluxes but steeper spectral indices.
To quantify the change in spectral index between the two regions, we calculated the average spectral index between the \textit{outer edge} and \textit{bright bar} points. We obtained $\langle {\alpha_{edge}} \rangle = -0.76 \pm 0.02$ and $\langle\alpha_{bar}\rangle = -0.94 \pm 0.02$, so the average spectral index is reduced by $-0.18$. The downstream region shows a lower flux density and spectral index than both the main shock and the filament.
A simple, uniform shock scenario cannot account for the observed trends and the observed spectral indices, which suggest a more complex physical picture. For example, a complex shock surface could accelerate particles through repeated DSA events throughout the radio relic, leading to flat spectral indices. Similarly, projection effects could blend freshly accelerated particles with the downstream medium along the LoS, originating flatter indices and influencing the morphology of the observed radio emission. At last, magnetic field inhomogeneities could enhance the radio emission in different regions of the relic.\\
It is believed that, due to their highly polarized radio emission, radio relics are mostly viewed edge-on \citep{Ennslin1998}, which would confine the merger axis to near the plane of the sky. Simulations \citep{Skillman2013} are also showing the same effect. Therefore, we would expect the projection effect in clusters hosting radio relics to be minimized. Despite that, local projection effect in the site of the radio relic could reproduce the observed ``bright bar'' feature \citep{Wittor2019}. Furthermore, \citet{Dominguez-Fernandez2024} showed that the Mach number distribution of the shock front plays the main role in shaping the observed surface brightness in radio relics. At last, the substructure could also be due to the adiabatic compression induced by a secondary shock, which increases the maximum synchrotron frequency emitted by a population of fossil electrons by a multiple of the compression factor, although the boosting which is usually observed in this case is moderate (factor of $\sim 2$), and the scenario has been ruled out for most cases of shocks \citep{Hoang2018}.\\
No distinct features were identified in the SW relic. Nevertheless, as discussed in Section 3, its spectral index remains flatter than expected from standard DSA models. This discrepancy may be attributed to projection effects or to a scenario in which the shock reached the SW region at a later stage, leading to spectral flattening.

\section{Summary and conclusions}\label{Summary}
We presented results from deep, wide-band uGMRT band 3 ($300-500\,$MHz) and band 4 ($550-750\,$MHz) radio observations of the galaxy cluster MACS J1752.0+4440. These observations were paired with archival JVLA L band ($1.6\,$GHz) data and LOFAR HBA data \citep[LoTSS-DR2]{Botteon2022} in order to perform a detailed spectral study of the cluster. Thanks to the high resolution of these observations, we were able to observe and characterize the features in the radio relics of MACS J1752. Our results are summarized as follows:
\begin{enumerate}
    \item The radio emission from the two radio relics closely follows a power-law, with integrated spectral indices of $\alpha_{int}^{NE} = -0.91 \pm 0.06$ and $\alpha_{int}^{SW} = -0.83 \pm 0.05$, with no signs of spectral steepening. These spectral indices are higher than what is normally expected from radio relics, and flatter than predicted by simple DSA, meaning that a more complex and realistic scenario must be considered.
    \item We detected a substructure along the NE relic, that we referred to as the ``bright bar'', throughout all of the radio bands. We observed no substructures along the SW relic.
    \item We produced brightness and spectral index profiles for both relics, showing how the NE relic peaks in surface brightness at the position of the ``bright bar'', while showing a secondary peak in spectral index at the same position. The SW relic instead has a brightness and spectral index peak close to the edge of the relic, followed by a sharp drop towards cluster center.
    \item Through a spectral-index curvature study, we showed how the spatial variation of the spectral index follows the substructures of the relics. Thanks to the color-color plot technique, we observed two spectra that are closely following a power-law, showing no signs of spectral curvature. The observed curvature suggests either localized re-acceleration, spatial mixing of multiple electron populations with different ages, or shock-induced compression of fossil plasma.
    \item We investigated the nature of the ``bright bar'' substructure. Following the hypothesis of a complex scenario, we investigated the distribution of spectral indices and fluxes of several points along the NE relic. This revealed two different radio emitting structures, represented by the ``outer edge'' and the ``bright bar'' substructures, suggesting a complex scenario.
\end{enumerate}

In summary, our results highlight the crucial role of high-resolution radio observations in unveiling the intricate substructures within radio relics. The detection of such substructures in this work and in other recent studies provides compelling observational evidence that challenges the validity of a simple scenario where a single shock front accelerates particles via DSA, instead pointing toward a more complex picture of particle acceleration and shock dynamics in the ICM.
Future studies of polarization, particularly through rotation measure synthesis, will be essential to probe the magnetic field orientation and to shed light on the physical origin of the observed substructures.

\begin{acknowledgements}

M. Della Chiesa acknowledges support from the ERC Consolidator Grant ULU 101086378.
A. Bonafede acknowledges support from the ERC-CoG $\vec B$ELOVED, GA n. 101169773.
LOFAR \citep{vanHaarlem2013} is the LOw Frequency ARray designed and constructed by ASTRON. It has observing, data processing, and data storage facilities in several countries, which are owned by various parties (each with their own funding sources), and are collectively operated by the ILT foundation under a joint scientific policy. The ILT resources have benefitted from the following recent major funding sources: CNRS-INSU, Observatoire de Paris and Université d’Orléans, France; BMBF, MIWF-NRW, MPG, Germany; Science Foundation Ireland (SFI), Department of Business, Enterprise and Innovation (DBEI), Ireland; NWO, The Netherlands; The Science and Technology Facilities Council, UK; Ministry of Science and Higher Education, Poland; Istituto Nazionale di Astrofisica (INAF), Italy. This research made use of the Dutch national e-infrastructure with support of the SURF Cooperative (e-infra 180169) and the LOFAR e-infra group, and of the LOFAR-IT computing infrastructure supported and operated by INAF, including the resources within the PLEIADI special “LOFAR” project by USCC of INAF, and by the Physics Dept. of Turin University (under the agreement with Consorzio Interuniversitario per la Fisica Spaziale) at the C3S Supercomputing Centre, Italy. The Jülich LOFAR Long Term Archive and the German LOFAR network are both coordinated and operated by the Jülich Supercomputing Centre (JSC), and computing resources on the supercomputer JUWELS at JSC were provided by the Gauss Centre for Supercomputing e.V. (grant CHTB00) through the John von Neumann Institute for Computing (NIC). This research made use of the University of Hertfordshire highperformance computing facility and the LOFAR-UK computing facility located at the University of Hertfordshire and supported by STFC [ST/P000096/1]. We thank the staff of the GMRT for making these observations possible. GMRT is run by the National Centre for Radio Astrophysics of the Tata Institute of Fundamental Research.
\end{acknowledgements}

\bibliography{bib}

\end{document}